\documentclass[%
 reprint,
 amsmath,amssymb,
 aps,
prb,
]{revtex4-2}

\usepackage{graphicx}
\usepackage{dcolumn}
\usepackage{booktabs}
\usepackage{bm}
\usepackage[english]{babel}
\usepackage{amsmath, graphicx, appendix}
\usepackage{color, xcolor}
\usepackage{natbib}

\begin{document}

\title{Intrinsic Wannier Functions for Hamiltonian downfolding}

\author{Shuoxue Li}
\email{sli7@caltech.edu}
 
\author{Garnet Kin-Lic Chan}%
\email{gkc1000@gmail.com}

\affiliation{Division of Chemistry and Chemical Engineering, California Institute of Technology, Pasadena, CA 91125, USA}
\affiliation{Marcus Center for Theoretical Chemistry, California Institute of Technology, Pasadena, CA 91125, USA}

\date{\today}

\begin{abstract}
Downfolding ab initio material band structure into a low-energy subspace spanned by orbitals of specified atomic character, a procedure known as Wannier downfolding, is a common task in the simulation of complex materials. We introduce the Intrinsic Wannier Function (IWF) method to Wannierize bands with given atomic character.  The method is non-iterative and requires only a single dimensionless parameter to disentangle bands. In benchmarks on silicon, graphene, and the three-band model of a mercury cuprate, we show that Intrinsic Wannier Functions provide high quality downfolded band structures compared to those from standard approaches such as Maximally Localized Wannier Functions, the Selected Columns of the Density Matrix method, and Projectability-Disentangled Wannier Functions. Further, their straightforward implementation and robustness position Intrinsic Wannier Functions as a general and useful tool for Wannier downfolding in materials electronic structure and in high-throughput applications.
\end{abstract}

\maketitle 

\section{Introduction}

Many computational studies of materials, particularly correlated materials~\cite{Plakida2010-un,Ramakrishnan2025-mq,Cui2025-kx,LaBollita2024-dm,Wang2025-qs,Kitatani2023-tq}, start by downfolding from an ab initio bandstructure to a model Hamiltonian~\cite{Miyake2009-uh,Hirayama2018-jw,Hirayama2022-au,Chang2024-hx}. This simplifies the complicated simulation of realistic materials while retaining elements of the specific material character relevant to the low-energy physics.

When generating the downfolded model, the key problem is how to choose the appropriate single-particle subspace to span. Typically, it is required that the downfolded space is spanned by a set of orbitals of specified local atomic character which yield bands that approximate the ab initio low-energy bands. This task is complicated by the fact that bands of different atomic character can mix, particularly in the higher-energy space. The most common method to do this utilizes Maximally Localized Wannier Functions (MLWF)~\cite{Marzari1997-dt,Souza2001-ut,Marzari2012-qh,Vitale2020-ce}, which performs a disentanglement procedure within an energy window and performs iterative minimization of the spread of the orbitals to obtain atomic-like functions. However, some disadvantages of this approach are the sensitivity of the resulting Wannier functions to the energy window, and the possibility of the spread minimization to yield multiple solutions. Thus other Wannierization procedures have also been proposed, for example, the selected columns of the density matrix (SCDM) method~\cite{Damle2015-en,Damle2018-bg}, which uses straightforward linear algebra to construct Wannier orbitals. However, without additional constraints, the SCDM procedure can yield Wannier functions that do not have recognizable atomic character~\cite{Qiao2023-hg}. The recently introduced projectability-disentangled Wannier function (PDWF) method~\cite{Qiao2023-hg} uses a pair of projectability thresholds, with optional energy windows, while keeping the local minimization problem of MLWF, and thus introduces even more complexity and parameters~\cite{Qiao2023-hg}.

In this work, we provide a simple alternative scheme, which we call the Intrinsic Wannier function (IWF) method. Motivated by the concept of Intrinsic Atomic Orbitals (IAO)~\cite{Knizia2013-sn,Cui2020-pf,Barison2022-kr,Zhu2024-kt}, the IWF construction creates localized orbitals without the need for any iterative protocol, unlike MLWF and PDWF. However, unlike SCDM, the input to IWF is a reference minimal basis of atomic orbitals, which ensures that the resulting Wannier functions retain good atomic character, and also allows for fine control of which bands one wishes to reproduce. Note that IWF are also distinct from IAO: IAO only exactly reproduce the occupied subspace and bandstructure, whereas, for an isolated set of bands, IWF reproduce the entire set. 
In the case where the bands are not isolated, to disentangle higher energy bands, we introduce a new type of smearing procedure based on the character of the bands. Unlike energy based smearing~\cite{Damle2018-bg}, the smearing parameter here is dimensionless, and is thus insensitive to the energy-scale of the band structure, and unlike the pair of projectability thresholds in PDWF and the optional energy window, there is only a single parameter. 

We organize the paper as follows: In Section~\ref{sec:theory} we describe the IWF construction.  Section~\ref{sec:results} then compares IWF to SCDM and MLWF downfolding for silicon and graphene and highlights its favorable characteristics. We then use IWF, MLWF and PDWF to construct a three-band downfolded model for a double-layer mercury cuprate. In this case, we find that the IWF downfolded bands are arguably more faithful to the original band structure than those from MLWF and comparable to those from PDWF, while not requiring an adjustment of an energy window or of projectability thresholds. Finally, we quantitatively compare the localization of the Wannier functions and the fidelity of the downfolded bands across all the methods. We present our conclusions and outlook in Section~\ref{sec:conclusion}.

\section{Theory\label{sec:theory}}

We start with a molecular setting before proceeding to the material band setting. Assume we have a computational basis $B_1 = \{| \mu \rangle \}$, and a reference minimal AO basis $B_2 = \{ | \tilde \mu \rangle \}$. Following the original IAO proposal, $B_2$ corresponds to the free atom AOs (although as also discussed in Ref.~\cite{Knizia2013-sn}, the final IAO properties are usually insensitive to the precise definition of `free atom AOs') and as a minimal basis, it is restricted to the valence and core AOs; we may also choose some subset of these to be $B_2$, e.g. only the valence AOs, or certain specific valence AOs. $B_1$, on the other hand, contains additional flexibility to describe the polarization of the AOs from the atomic environment. $B_1$ might be a large AO basis, but in fact any flexible basis, such as a plane-wave basis, suffices. 

Now choose a set of (occupied and virtual) molecular orbitals (MO) $\{ |i\rangle \}$, expanded in $B_1$. Our goal is to obtain an atomic-like basis which spans the space of chosen $\{| i\rangle\}$ (and vice versa), thus the number of chosen $\{ |i \rangle \}$ is taken to be equal to the dimension of $B_2$. 
We first define the MO projector $O = \sum_i | i \rangle \langle i |$. Then the projected basis, $\mathrm{orth}\{ O |\tilde \mu\rangle \}$ spans $ \{ |i\rangle \}$, so long as it is full rank.

\begin{figure*}[htbp]
    \centering
    \includegraphics[width=0.83\linewidth]{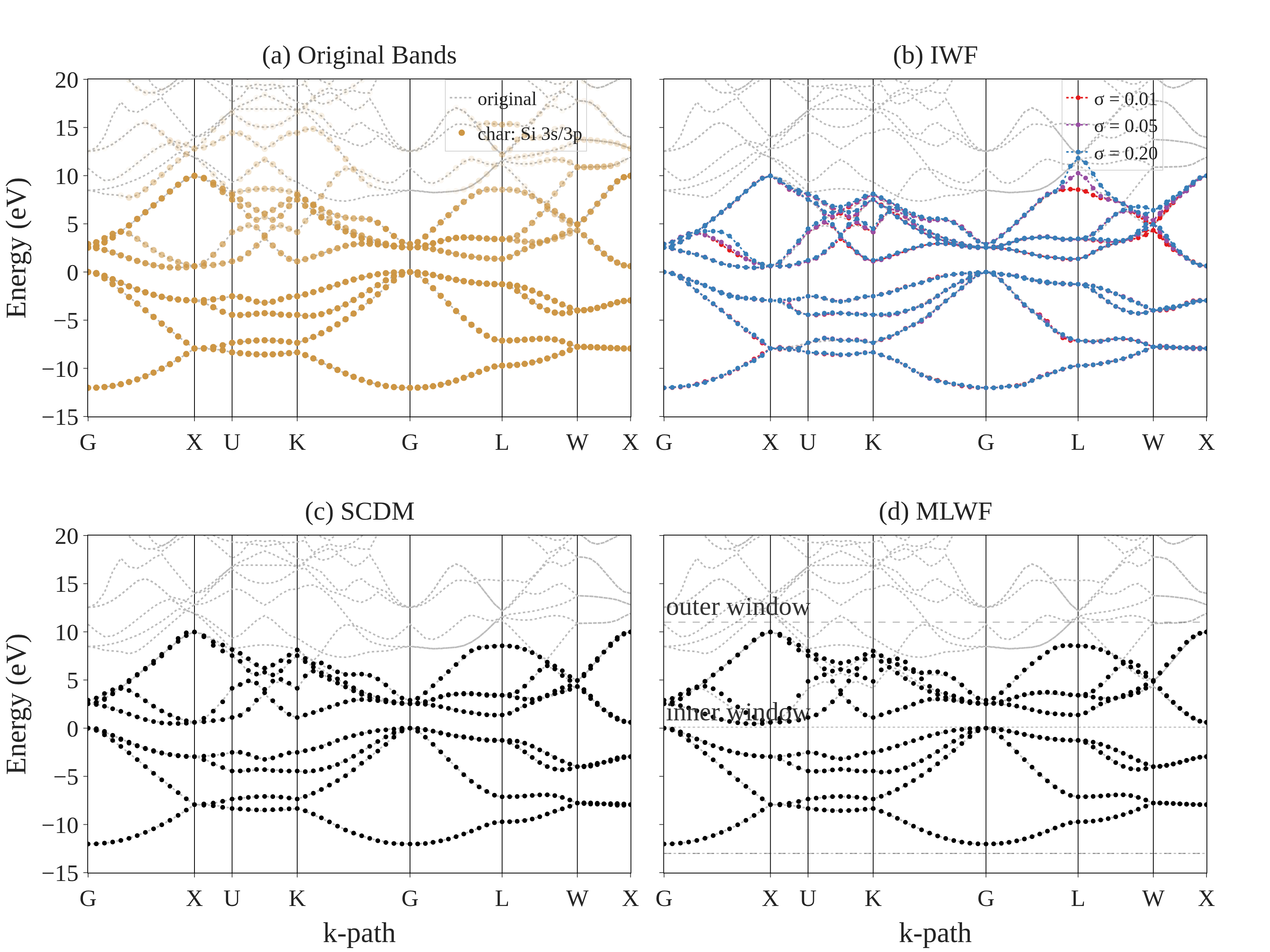}
    \caption{Comparison of FCC silicon band structure with different localization methods. (a) Original band structure showing the (total) Si 3s/3p character of the bands. (b) Intrinsic Wannier Function downfolding to the 3s/3p bands with different character smearings $\sigma$. (c) SCDM downfolding with chemical potential $\mu = 5.0\,\text{eV}$ and the energy smearing parameter $\sigma = 2.0\,\text{eV}$; (d) MLWF downfolding with inner energy window $[-13.0\,\text{eV}, 0.1\,\text{eV}]$ and outer energy window $[-13.0\,\text{eV}, 11.0\,\text{eV}]$.}
    \label{fig:Si}
\end{figure*}

We extend this idea to construct intrinsic Wannier functions. The molecular orbitals become bands that carry crystal momentum labels, $|i^\mathbf{k}\rangle$, and the $B_2$ basis is a crystalline/Bloch AO basis, 
 \begin{equation}
     |\tilde \mu\rangle \to |\tilde \mu^\mathbf{k}\rangle = \sum_\mathbf{T} |\tilde \mu(\mathbf{r}-\mathbf{T})\rangle e^{i \mathbf{k}\cdot \mathbf{T}}  
     \end{equation}
We first assume that the bands are well separated energetically from all the other bands, and have well-defined atomic characters. 
Then we construct the IWFs at each $\mathbf{k}$-point as 
\begin{align}
    \{ |\mu^{\text{IWF}, \mathbf{k}}\rangle \} = \text{orth} \{ O |\tilde{\mu}^\mathbf{k}\rangle \}
\end{align}
where $\text{orth}\{\cdots\}$ means the L\"owdin orthogonalization of a basis set.

Assuming a standard Monkhorst-Pack grid with $N_\mathbf{k}$ points, the IWFs in real space in the primitive cell are obtained as
\begin{equation}
    \{ |\mu^\text{IWF} \rangle \} = \frac{1}{N_\mathbf{k}} \sum_\mathbf{k} |\mu^{\text{IWF}, \mathbf{k}}\rangle
\end{equation}
The IWFs obtained this way clearly span the chosen crystalline bands. In particular, given the one-particle Hamiltonian $\hat{h}$ and the projector into the IWF space, $\hat{P}^\text{IWF}$, then diagonalizing $\hat P^\text{IWF} \hat{h} \hat P^\text{IWF}$ will exactly reproduce the band-structure of the selected bands. 

If the bands to be downfolded are not well separated, the physical goal is not to exactly reproduce the band-structure of a fixed selected set of bands. This is because there are band crossings (typically in the conduction bands) that appear in the large basis $B_1$ which would not appear in the basis $B_2$, a phenomenon known as entanglement of the band structure~\cite{Souza2001-ut}. These band crossings lead to the given atomic characters being spread out across different numbers of conduction bands at different $\mathbf{k}$ points. The physical goal is to recover a `diabatic basis', i.e. the band structure that would appear in a valence basis without the crossings induced by the larger basis. To obtain such a diabatic picture, it is necessary to consider the bands that are mixed together due to the crossing.

There is no unique prescription for constructing the diabatic basis, but out of the different disentanglement schemes in the literature, one of the simplest is to apply a smearing function to select an average over the entangled bands, as originally described in the SCDM method~\cite{Damle2018-bg}. Here, in keeping with the character based selection of the target bands, rather than smearing in energy space we apply a character based smearing function. We define the character of the bands $i$ as
$\chi_{i}^{\mathbf{k}} = \sum_{\mu'} |\langle i^\mathbf{k}|\mu^{\prime \mathbf{k}} \rangle|^2 \in [0, 1]$, which requires a set of orthogonal localized crystalline atomic orbitals $\{|\mu^{\prime \mathbf{k}}\rangle\}$. If $B_2$ is the only atomic-like basis in the calculation (e.g. when using a plane-wave computational basis) then $|\mu^{\prime \mathbf{k}}\rangle$ can be taken to be the L\"owdin orthogonalized set of $B_2$. If $B_1$ is also an atomic basis, then we can evaluate the character alternatively in the $B_1$ basis after suitable orthogonalization, e.g. $|\mu^{\prime \mathbf{k}}\rangle$ is obtained by the meta-L\"owdin procedure (separate L\"owdin orthogonalization by core, valence, and virtual AO blocks~\cite{Sun2014-nq}). We will see below that the choice does not make much difference. Then
\begin{align}
    \hat{O}^\mathbf{k} = \sum_i f(\chi_i^{\mathbf{k}}) |i^\mathbf{k}\rangle \langle i^\mathbf{k}|
    \label{eq:mo-projector}
\end{align}
where
\begin{align}
    f(x) = \dfrac{\text{FD} (-x) - \text{FD}(0)}{\text{FD}(-1) - \text{FD}(0)}
\end{align}
and FD is a Fermi-Dirac function defined as
\begin{equation}
    \mathrm{FD}(x) = \dfrac{1}{1 + \exp\left(\dfrac{x - \mu}{\sigma}\right)}
\end{equation}

Note that the above shifted and scaled Fermi-Dirac-like function satisfies $f(0) = 0$ and $f(1) = 1$. $\mu$, analogous to the chemical potential, is adjusted so that $\sum_i f(\chi_i^{\mathbf k}) = N$, the size of the $B_2$ basis, which is the total number of desired bands. The well-separated band limit is obtained by taking $\sigma \to 0$, which sets $f(\chi_i^{\mathbf k}) = 1$ only for the bands of maximal atomic character of interest (and otherwise $0$). Non-zero $\sigma$ gives the typical range of atomic characters that enter into the smeared projector $\hat{O}$, and we refer to it as the smearing parameter, although unlike an energy smearing parameter, it is dimensionless.

If the $\mathbf{k}$-mesh samples a $\mathbf{k}$-point where there exist orbitals that are exactly degenerate in energy, the construction of the projector in Eq.~\ref{eq:mo-projector} depends on the gauge in the degenerate space. Although such points are a vanishing set and thus this problem is uncommon in practice, we can simply fix the gauge. For example, diagonalizing the character operator
\begin{equation}
    \hat \chi^{\mathbf{k}} = \sum_{\mu'} (| \mu^{\prime\mathbf{k}} \rangle \langle \mu^{\prime \mathbf{k}} |)
    \label{eq:scaled-char-mat}
\end{equation}
in each degenerate space and using the eigenvalues and eigenvectors as $\chi_i$ and $| i^{\mathbf{k}} \rangle$ respectively is one choice. This, however, leads to a discontinuity in the band coefficients around the degenerate $\mathbf{k}$-point. This can be resolved by instead diagonalizing the scaled character matrix for all bands at every $\mathbf{k}$-point,
\begin{equation}
    \tilde \chi_{ij}^{\mathbf{k}} = \exp\left(-\dfrac{|\varepsilon_{i}^{\mathbf{k}} - \varepsilon_{j}^{\mathbf{k}}|^2}{2 \varepsilon_0^2 }\right) \langle i^{\mathbf{k}} | \hat \chi^{\mathbf{k}} | j^{\mathbf{k}} \rangle
\end{equation}
where $\varepsilon_i^\mathbf{k}$, $\varepsilon_j^\mathbf{k}$ are the band eigenvalues, and $\varepsilon_0$ is a smoothing coefficient with the unit of energy. We then substitute $\chi_i$ and $| i^{\mathbf{k}} \rangle$ in Eq.~\ref{eq:mo-projector} with the eigenvalues and the eigenvectors of this matrix. One sees that this smoothly reduces to the original prescription in Eq.~\ref{eq:mo-projector} away from any degenerate point.

The IWF algorithm thus consists of the following steps:
\begin{itemize}
    \item We start with the initial $B_2$ basis set of appropriate AO character. 

    \item For well-separated bands, we pick the $N$ bands with largest character and apply the IWF construction, while for entangled bands, we use character based smearing with a finite spread parameter $\sigma$.
    \item The downfolded band structure then either precisely agrees with the chosen bands (for well-separated bands) or is a diabatic approximation to it. 
    At an arbitrary $\mathbf{k}$-point, the band structure can either be obtained by constructing the IWF projected Hamiltonian $\hat{P}^\mathbf{k} \hat{h} \hat{P}^\mathbf{k}$ and diagonalizing, or alternatively, the IWF projected Hamiltonian can be constructed at a pre-defined set of $\mathbf{k}$-points (e.g. a Monkhorst-Pack grid) and the eigenvalues at an arbitrary $\mathbf{k}$ point computed by Wannier interpolation.
\end{itemize}

\section{Results and Discussion\label{sec:results}}

\subsection{Computational Details}

The ab initio multigrid DFT calculations~\cite{Beck2009-it} were performed using the PySCF quantum chemistry software package~\cite{Sun2018-wq,Sun2020-kp,Sun2026-bw}, using the GTH-cc-pVDZ basis set~\cite{Ye2022-hv,Cui2025-kx}, the GTH-Pade pseudopotential~\cite{VandeVondele2007-yz}, and the local-density approximation (LDA) functional~\cite{Kohn1965-wr,Sahni1988-ov}. $8 \times 8 \times 8$, $10 \times 10 \times 1$ and $8 \times 8 \times 2$ Monkhorst-Pack $\mathbf{k}$-point grids were used for silicon, graphene and $\text{HgBa}_2\text{CaCu}_2\text{O}_6$, respectively. The convergence with respect to the $k$-mesh for $\text{HgBa}_2\text{CaCu}_2\text{O}_6$ is validated by checking the dependence of the IWF bands and one-body properties on the mesh; see Fig.~\ref{fig:hbcco_kconv} in the Supplementary Information for details. The minimal $B_2$ basis used to construct the IAOs was generated starting from the extracted minimal basis part of the cc-pVTZ basis~\cite{Dunning1989-yy,Dunning2001-qh} that corresponds to the \texttt{minao} basis in PySCF, followed by the spherically averaged atomic Hartree-Fock calculation to redetermine the basis contraction coefficients. Except where indicated otherwise, we evaluated atomic character using meta-L\"owdin orthogonalized atomic orbitals in the computational (GTH-cc-pVDZ) basis~\cite{Sun2014-nq} and use the simple (non-degenerate band) definition of the projector in Eq.~\ref{eq:mo-projector} rather than the scaled character matrix.

For comparison to other approaches, we obtained MLWF and PDWF results through a Python interface to the Wannier90 code~\cite{Mostofi2008-to,Mostofi2014-io,Pizzi2020-aq} automatically generated with the f90wrap tool~\cite{Kermode2020-st}, with the initial trial state chosen as hydrogenic wavefunctions as used in traditional MLWF procedures~\cite{Marzari1997-dt}. We also compare to the SCDM method using the $\mathrm{erfc}$ smearing function with parameter $\sigma = 2.0\,\text{eV}$ and Becke grids~\cite{Becke1988-xx}. As discussed below, we used both an inner energy window (in which the band structure is exactly reproduced) and an outer energy window in the MLWF procedure. The SCDM implementations were modified from the versions in the LibDMET package~\cite{Cui2020-nf,Cui2022-yg,Cui2025-kx}. For more details of the MLWF, PDWF and SCDM procedures, we refer to Refs.~\cite{Marzari1997-dt,Souza2001-ut}, Ref.~\cite{Qiao2023-hg}, and Refs.~\cite{Damle2015-en,Damle2018-bg}, respectively. For all methods, we computed the downfolded band-structure by Wannier interpolation using the Wannier functions obtained on the Monkhorst-Pack mesh.

\subsection{Silicon}

We start by discussing the $3s$ and $3p$ bands of face-centered cubic (FCC) silicon. The total $3s$/$3p$ character of the bands is shown in Fig.~\ref{fig:Si}(a). We clearly see four disentangled, well-separated bonding bands below the Fermi level, as well as four anti-bonding bands that are entangled with other virtual bands. 

In Figs.~\ref{fig:Si}(b),(c),(d), we show the downfolded band structure from IWF, SCDM, and MLWF. In all cases, the filled bands are exactly reproduced, which is guaranteed by the full occupancy of these bands as measured by the IWF character ($f(\chi_i) \sim 1$), full occupancy in the SCDM Fermi-Dirac function, and the use of the inner energy window to contain these bands in the MLWF. However, the behavior of the four conduction bands varies slightly amongst the three methods. Some visible differences are between the U and K points and the L and W points, where there is significant entanglement with higher bands. 

\begin{figure*}
    \centering
    \includegraphics[width=0.70\linewidth]{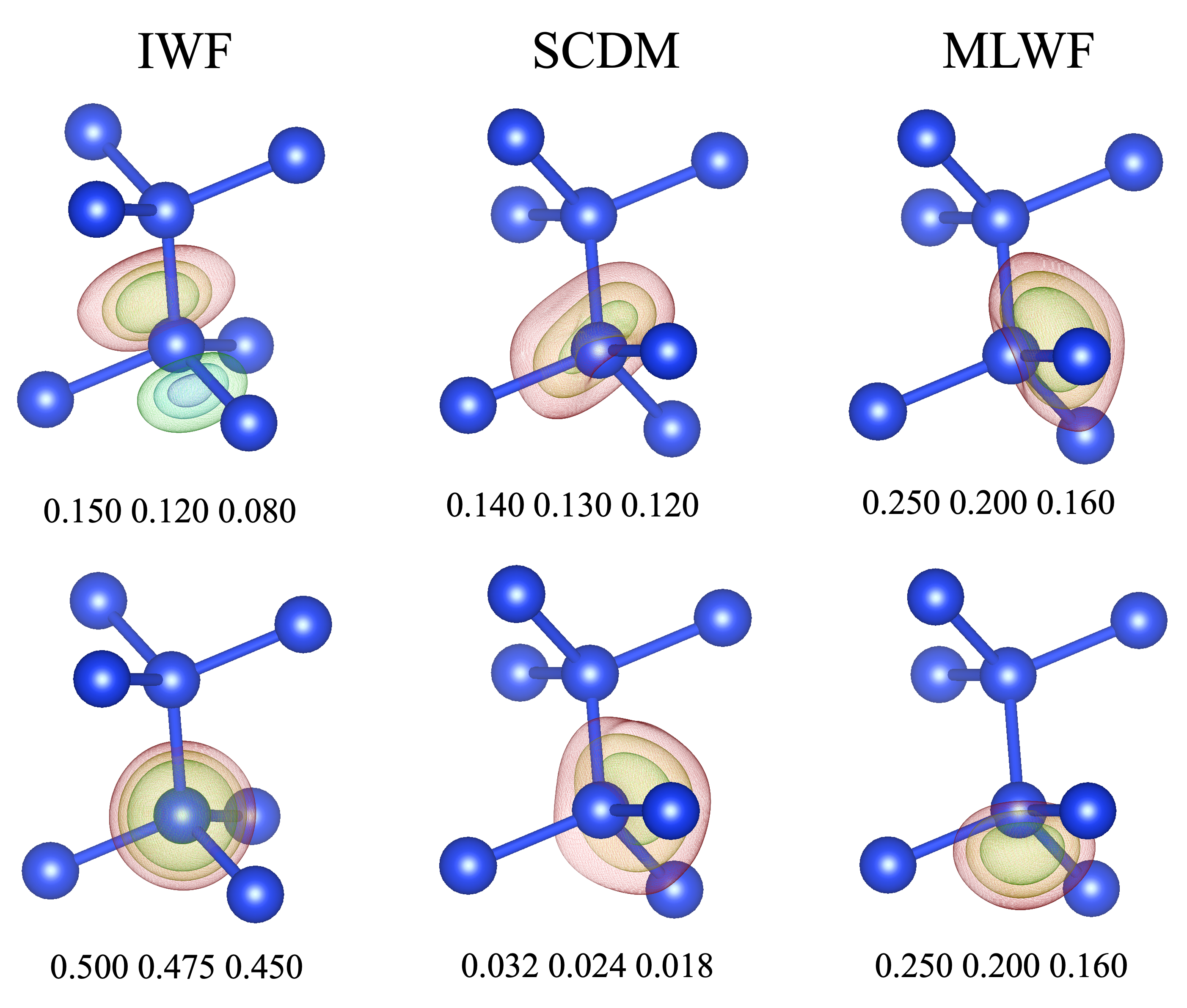}
    \caption{Selected silicon Wannier function orbitals obtained from IWF ($\sigma = 0.05$), SCDM and MLWF. The three numbers under each orbital are the isosurface values.}
    \label{fig:SiOrbitals}
\end{figure*}

In Fig.~\ref{fig:SiOrbitals}, we plot the Wannier functions obtained from IWF, SCDM, and MLWF. The IWF functions retain the AO character of the $B_2$ basis, and thus clearly resemble $3s$ and $3p$ orbitals. In contrast, the SCDM orbitals are irregular in their shapes and amplitudes, indicating that they have mixed $3s$ and $3p$ character without a clearly interpretable meaning. The MLWF functions depend on the initial guess and in Si there are multiple minima in the objective function~\cite{Vitale2020-ce}; the ones we find are in the form of $sp^3$ `back-bonding' hybrids~\cite{Vitale2020-ce}, although we started the projection from the individual $3s$ and $3p$ orbitals.

In Fig.~\ref{fig:Si}(b) we show the effect of changing the smearing parameter (between 0.01 and 0.20) on the IWF band structure. Near the X and L points, the larger smearing leads to some mixing with higher bands (which contain $3s$/$3p$ character), slightly raising the energy of the IWF bands near those points. 

\subsection{Graphene}

\begin{figure*}[htbp]
    \centering
    \includegraphics[width=0.85\linewidth]{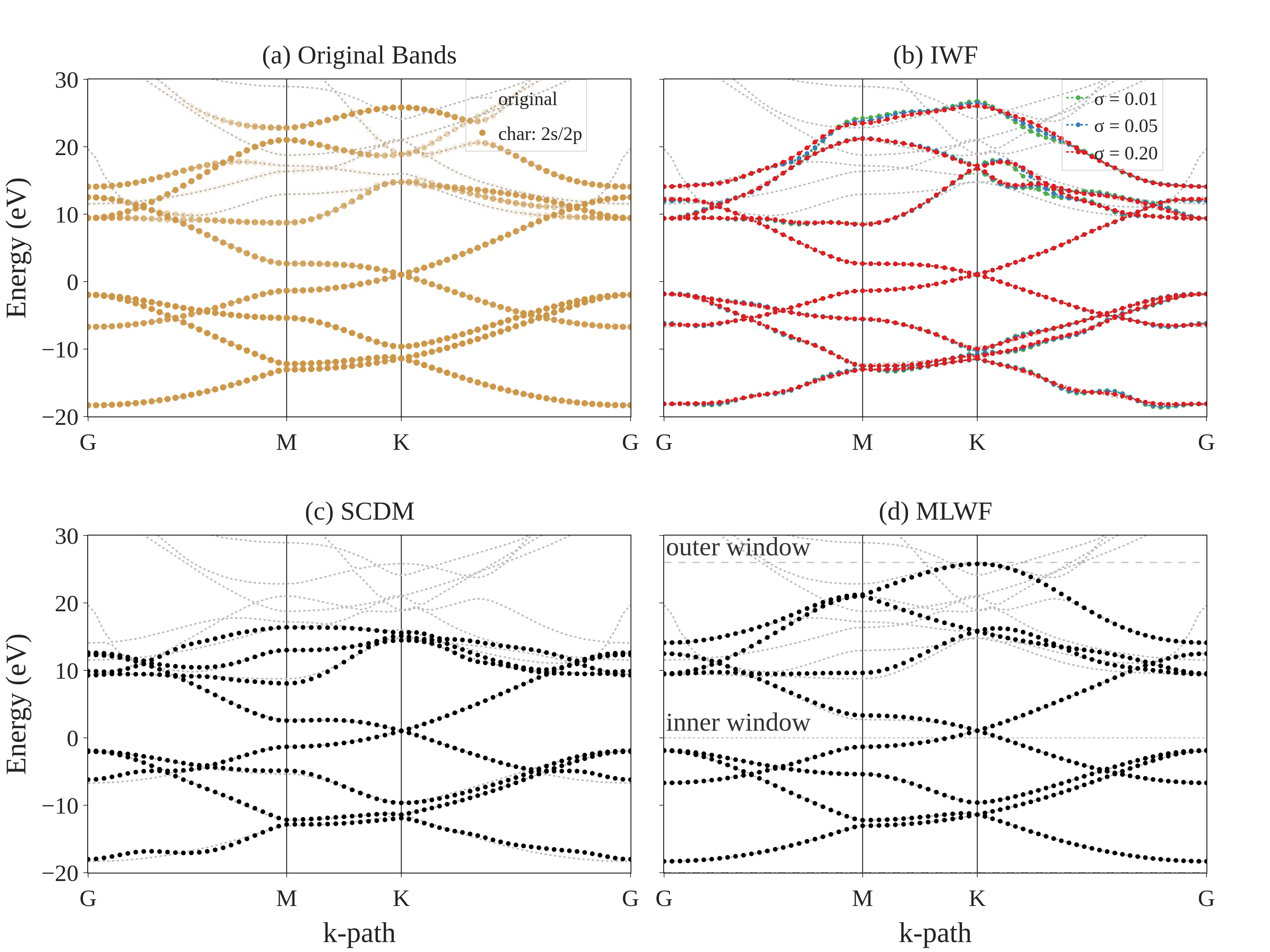}
    \caption{Comparison of graphene band structure with different localization methods. (a) Original bands showing the (total) C $2s$/$2p$ character. (b) Intrinsic Wannier Function downfolding to the $2s$/$2p$ bands. (c) SCDM downfolding with $\mu = 11.0\,\text{eV}, \sigma = 2.0\,\text{eV}$. (d) MLWF downfolding with  inner energy window [-20.0\,eV, 0] and outer energy window [-20.0\,eV, 26.0\,eV].}
    \label{fig:Graphene}
\end{figure*}

In Fig.~\ref{fig:Graphene} we show the $2s$ and $2p$ bands of a 2D graphene sheet.  Note that in energy space, these bands are situated above and below the Dirac point, and the virtual bands largely entangle with the free-electron bands in the energy space.

The downfolded band structures from IWF and MLWF are quite similar: both yield smooth bands that are identical to the original bands below the Fermi level and which provide a similar and reasonable disentangling of the conduction bands. There is a slight difference between the methods for the two highest bands near the M point: In MLWF they are nearly degenerate (note this degeneracy differs even amongst different MLWF methods as discussed in Ref.~\cite{Qiao2023-hg}) while when using IWF, the separation of the two bands that corresponds to the original band structure is conserved regardless of the smearing parameter $\sigma$. SCDM on the other hand gives a very different picture of the conduction bands, which do not follow the target band structure. This is because the SCDM procedure does not make use of the atomic character of the bands: the columns selected by the column-pivoted QR factorization (QRCP) are taken from a density matrix whose occupations are set by the energy smearing function, so that, up to the smearing, SCDM reproduces a given number of bands below the chosen chemical potential $\mu$ irrespective of their atomic character. If $\mu$ is placed at the top of the MLWF outer window in order to reach the highest bands of $2s$/$2p$ character, more than the 8 target bands lie below $\mu$, and SCDM fails to reproduce some of the occupied bands; if $\mu$ is chosen lower (e.g. at 11~eV here), it reproduces the bands below 11~eV well, but not those above. In IWF, the target bands are instead identified directly by the projection criterion on their atomic character, which avoids this problem without an additional fitting step.

\begin{figure*}[htbp]
    \centering
    \includegraphics[width=0.95\linewidth]{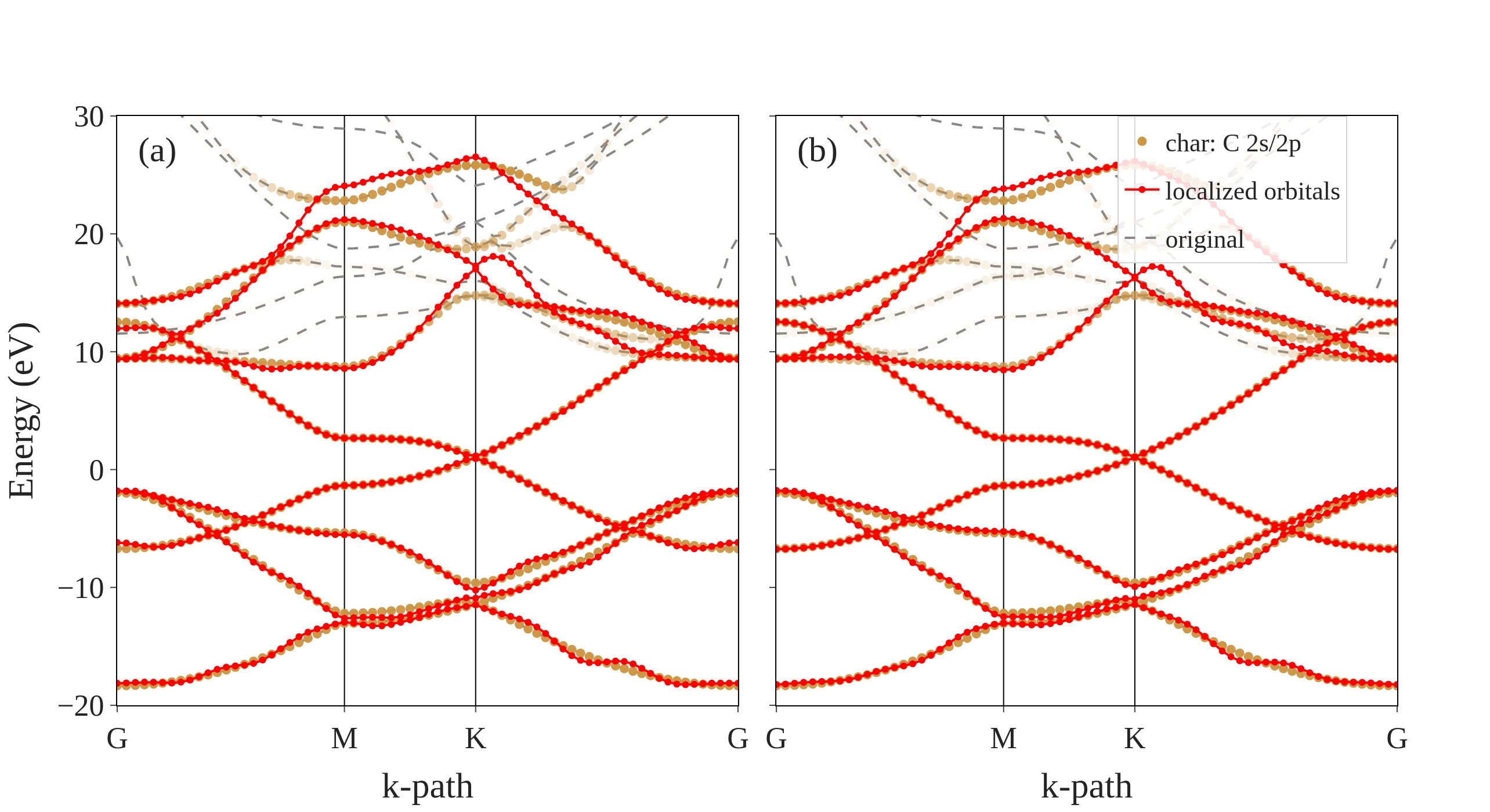}
    \caption{Variants of the IWF method using graphene as a benchmark. (a) IWF using the minimal $B_2$ basis to calculate the character instead of the meta-L\"owdin orthogonalized $B_1$ basis; (b) IWF using the scaled character matrix with parameter $\varepsilon_0$ set to 1.0~eV. The IWF smearing parameter $\sigma = 0.05$ for both cases.}
    \label{fig:graphene-more-discussions}
\end{figure*}

In Fig.~\ref{fig:graphene-more-discussions}(a) we show the effect of changing the basis to evaluate the character of the bands from the GTH-cc-pVDZ basis to the minimal basis: we see the effect is small and the downfolded band structures are visually indistinguishable. Similarly, we show the effect of using the scaled character matrix with $\varepsilon_0=1.0$~eV in Fig.~\ref{fig:graphene-more-discussions}(b). Again, this leads to little difference in the downfolded band structure.

\subsection{Three-band downfolding of the double-layer cuprate $\text{HgBa}_2\text{CaCu}_2\text{O}_6$}

\begin{figure}[htbp]
    \centering
    \includegraphics[width=0.95\linewidth]{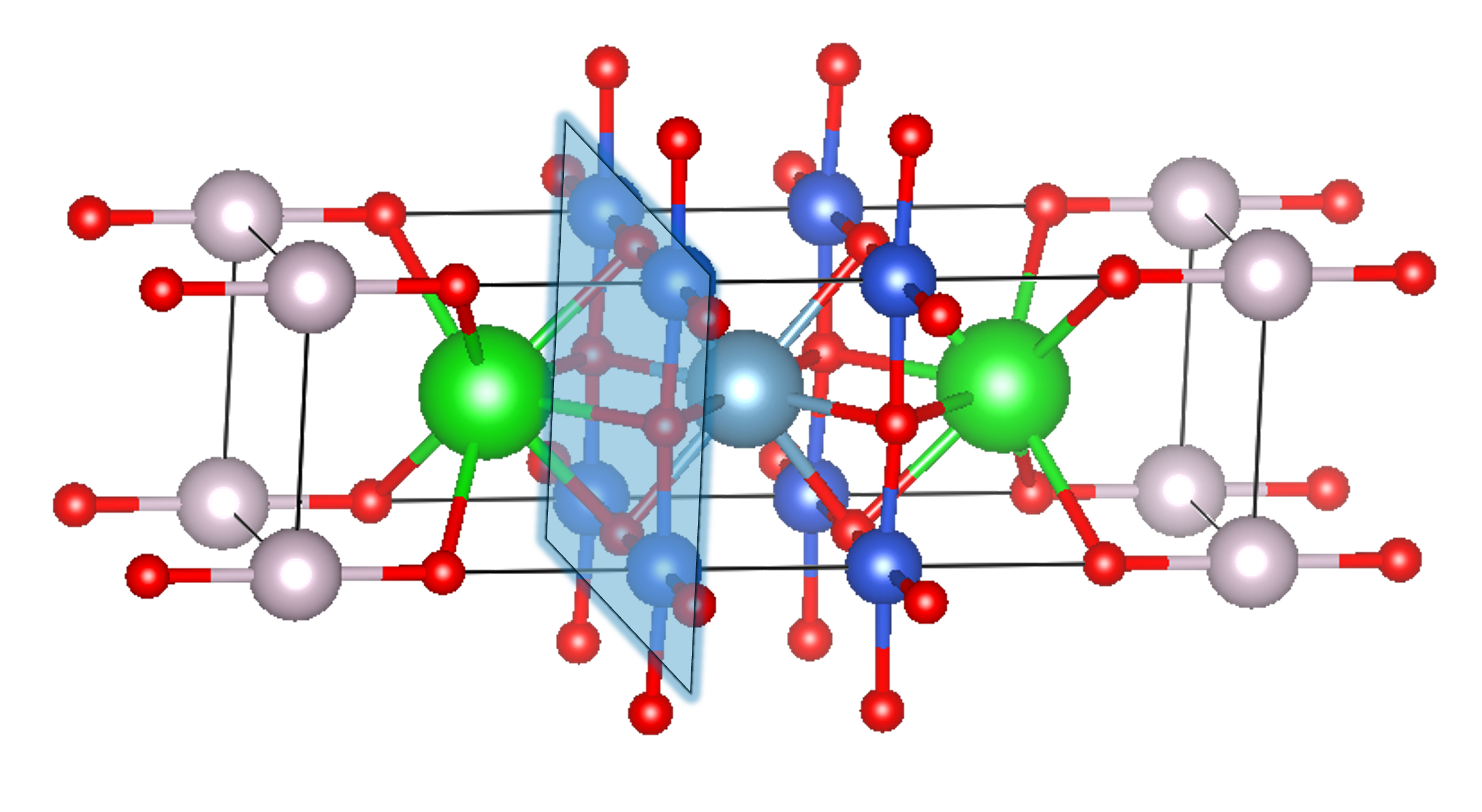}
    \caption{Illustration of the crystal structure and the CuO$_2$ plane for the material $\text{HgBa}_2\text{CaCu}_2\text{O}_6$.}
    \label{fig:cuprate-structure}
\end{figure}

\begin{figure*}[htbp]
    \centering
    \includegraphics[width=1.00\linewidth]{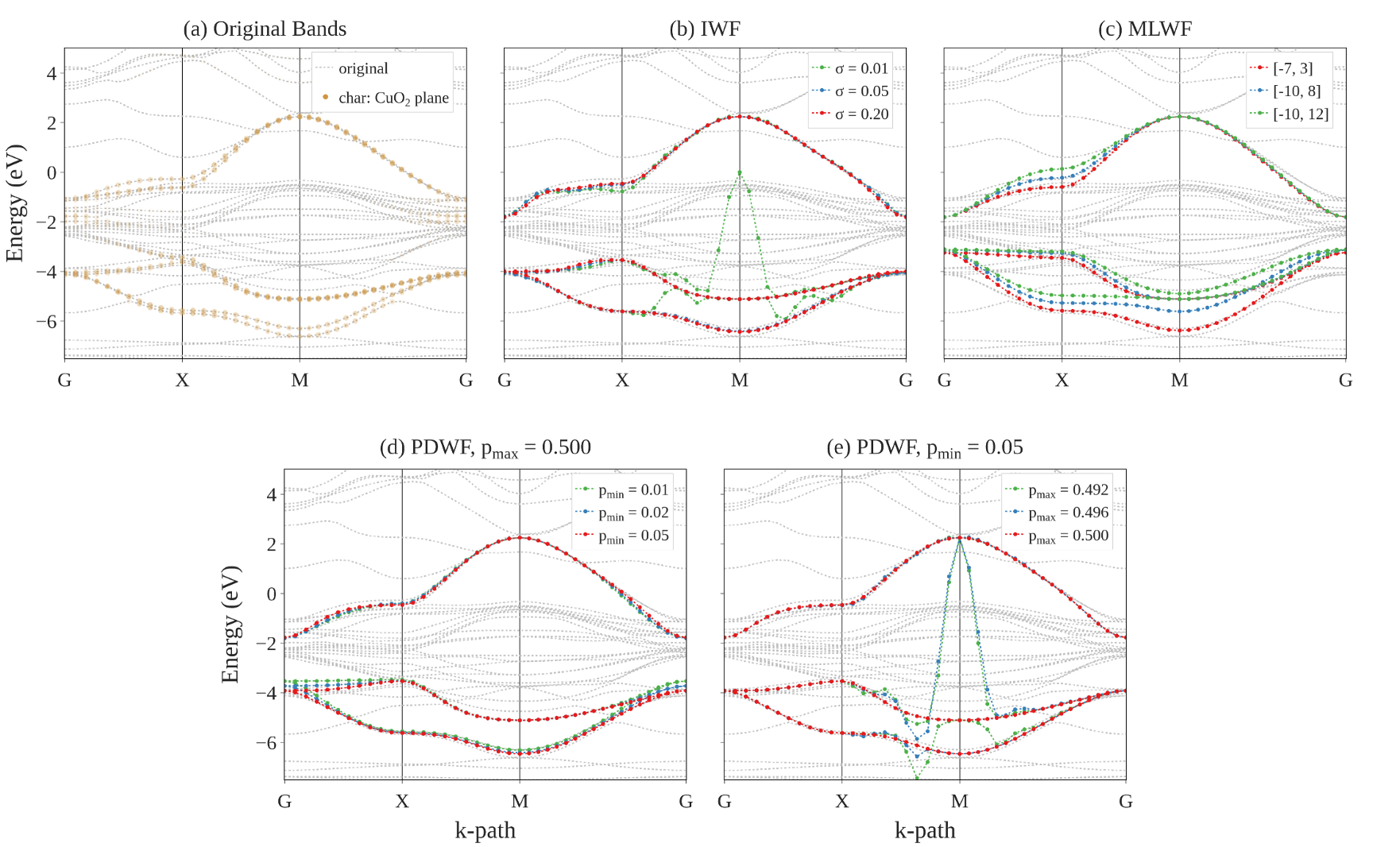}
     \caption{Comparison of the $\text{HgBa}_2\text{CaCu}_2\text{O}_6$ band structure with different localization methods. (a) Original band structure and total character in the single CuO\textsubscript{2} plane. (b) IWF downfolded bands with different smearing parameters $\sigma$; (c) MLWF downfolded bands with different ranges of the outer energy window (shown in the legend, unit: eV). The inner energy window is not set here since the three-band structure is not easily disentangled from other bands; (d) PDWF downfolded bands with $p_{\max} = 0.50$ and different lower projectability thresholds $p_{\min}$; (e) PDWF downfolded bands with $p_{\min} = 0.05$ and different upper projectability thresholds $p_{\max}$.}
    \label{fig:HBCCO}
\end{figure*}

 We finally illustrate the use of IWF to downfold a complicated cuprate band structure in order to derive a three-band model of a CuO\textsubscript{2} plane, i.e., consisting of the $3d_{x^2-y^2}$ of copper, $2p_x$ oxygen orbital for the Cu-O bond along the $x$-axis, and the $2p_y$ oxygen orbital for the Cu-O bond along the $y$-axis. We consider the material $\text{HgBa}_2\text{CaCu}_2\text{O}_6$  whose crystal structure is illustrated in Fig.~\ref{fig:cuprate-structure}, and which has two layers of CuO\textsubscript{2} planes. The band structure is shown in Fig.~\ref{fig:HBCCO}(a), showing roughly three near-degenerate band groups, where the near-degeneracy appears due to the double layer structure. 

As our goal is to obtain a 3-band model for a single plane, we use only a minimal three-band basis on a single copper plane for the IWF, and similarly in the MLWF and PDWF we select the Cu $d_{x^2-y^2}$ and O $p_x$, $p_y$ projections for 3 target atomic positions on a single layer. We compare the band structure of the projected Hamiltonians in Fig.~\ref{fig:HBCCO} and also compare the on-site and tight-binding Hamiltonian parameters in Table~\ref{tab:hbcco_hoppings} of the Supplementary Information.

In Fig.~\ref{fig:HBCCO}(b) we show the downfolded band structures from the IWF for different smearing parameters $\sigma$. When $\sigma = 0.01$ (small smearing) the downfolded band energies jump around between the X and M points. This occurs because selecting the 3 largest projections selects a subset of the 6 bands, and without smearing, different subsets are selected at different $\mathbf{k}$ points. However, as we increase $\sigma$ to 0.05 or 0.20, we mix in sufficient representation from all 6 bands that we recover a smooth model band structure.

The MLWF results for various outer energy windows are shown in Fig.~\ref{fig:HBCCO}(c). When the bounds of the energy window are near the extrema of the three degenerate groups ([-7~eV, 3~eV]), we find a better alignment of the band structure than for larger energy windows ([-10~eV, 8~eV] and [-10~eV, 12~eV]). Nevertheless, none of the energy window choices recovers the correct ab initio band structure for the lowest two degenerate groups between the G and X points and between the M and G points. We also compare to the recently proposed projectability-disentangled Wannier functions (PDWF)~\cite{Qiao2023-hg} in Fig.~\ref{fig:HBCCO}(d),(e). In PDWF, an energy window is optional; instead, Bloch states with projectability onto the target atomic orbitals above $p_{\max}$ are kept, those below $p_{\min}$ are discarded, and the remaining states enter the disentangling procedure. If we set $p_{\max}$ to be less than 0.5, we find some unphysical behavior in the model band structure, since band groups with character $\ge 0.5$ are fully represented by the Wannier functions. Setting $p_{\max}=0.5$, where all six important bands enter the disentangling procedure, we obtain a qualitatively correct band structure, but we need to increase $p_{\min}$ from zero for the PDWF bands to capture the correct behavior at the G point. IWF with a tiny smearing parameter has similar artifacts to PDWF with too small $p_{\max}$, but for a moderate smearing recovers a reasonable band structure, without requiring a fine-tuned energy window (as in MLWF). 

\begin{figure*}[htbp]
    \centering
    \includegraphics[width=0.80\linewidth]{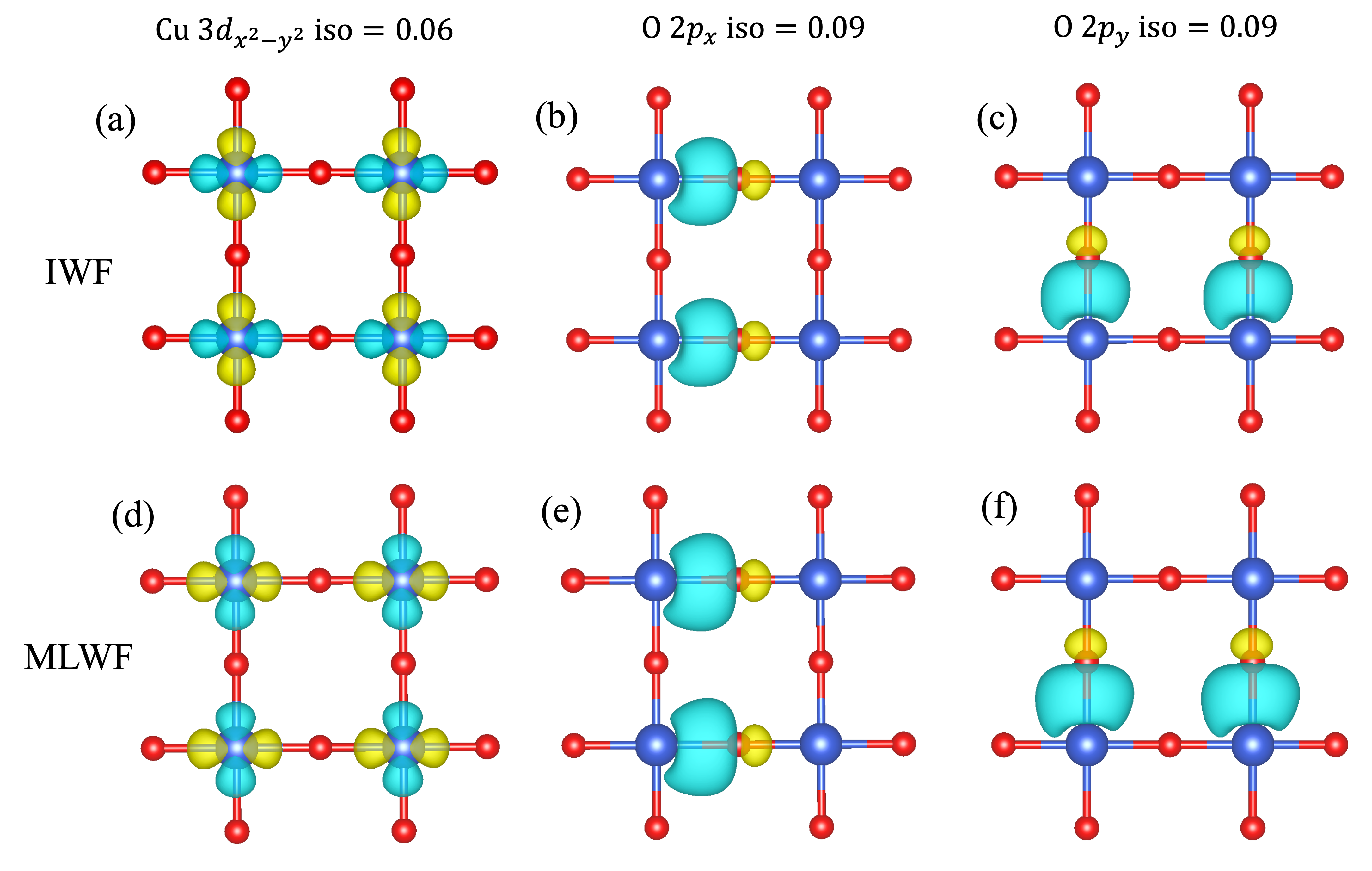}
    \caption{Shapes of the orbitals generated from IWF ((a)--(c), with $\sigma = 0.05$) and MLWF ((d)--(f), with outer energy window [-7~eV, 3~eV]). ``iso'' denotes the value of the orbital isosurface.}
    \label{fig:cuprate-orbs}
\end{figure*}

Fig.~\ref{fig:cuprate-orbs} shows the three-band orbital shapes of the IWFs compared with the MLWFs. All of them show the correct atomic character, including the symmetric relation between the polarized O $2p_x$ and O $2p_y$ orbitals. This shows that the IWF method can yield faithful and improved band structure without sacrificing the interpretable spatial atomic character.

\subsection{Quantitative analysis of the methods}

To quantify the differences between IWF and the other localization schemes, we report three metrics for all the systems studied above in Table~\ref{tab:spread_metrics}. The first two measure the spatial localization of the Wannier functions: the band-diagonal spread $\Omega_\mathrm{D}$ in the decomposition of the Marzari-Vanderbilt spread functional (Eq.~20 of Ref.~\cite{Marzari1997-dt}), and the largest individual spread $\max_n (\langle r^2\rangle_n - \bar{\mathbf{r}}_n^2)$ amongst the Wannier functions. The third measures the fidelity of the downfolded band structure, and we define it as the band energy spread
\begin{equation}
    S = \dfrac{1}{N_\mathbf{k}} \sum_\mathbf{k} \sqrt{\dfrac{1}{N}  \sum_{iI} \left(\varepsilon_{i}^{\mathbf{k}} - \tilde\varepsilon_{I}^{\mathbf{k}}\right)^2 \left| \langle i^{\mathbf{k}} | \tilde I^{\mathbf{k}} \rangle \right|^2 }
    \label{eq:energy-spread}
\end{equation}
where $|i^\mathbf{k}\rangle$ and $\varepsilon_i^\mathbf{k}$ are the ab initio bands and band energies, $|\tilde I^\mathbf{k}\rangle$ and $\tilde\varepsilon_I^\mathbf{k}$ are the eigenvectors and eigenvalues of the Hamiltonian projected onto the Wannier functions, and $N$ is the number of Wannier functions. Compared to the average band distance used in Ref.~\cite{Qiao2023-hg}, $S$ weights each ab initio band by its overlap with the downfolded eigenvectors, and thus does not require an explicit assignment of the ab initio bands to the downfolded bands.

Several general trends can be seen in Table~\ref{tab:spread_metrics}. First, the spatial spreads are ordered as MLWF $<$ IWF $<$ SCDM in all systems, and the real-space Hamiltonian decay rates in Table~\ref{tab:hbcco_hamiltonian_decay} of the Supplementary Information also give MLWF $>$ IWF. This is expected, since MLWF explicitly minimizes the spread functional while IWF does not, and SCDM does not constrain the atomic character of the orbitals. Second, the IWF orbitals become more localized as the smearing parameter $\sigma$ increases, which is due to the increasing mixture with more orbitals: in the limit that all bands are equally mixed in IWF, we recover the (orthogonalized) AO basis. Third, the band energy spread $S$ of IWF and SCDM is generally smaller than that of MLWF. This is because they are constructed from the selected bands themselves, through the character-based projector in IWF and through the column-pivoted QR factorization of the density matrix in SCDM, whereas MLWF does not take the band energies into account outside the inner energy window. Finally, whenever the downfolded band structure is smooth (i.e. excluding IWF with $\sigma = 0.01$ for $\text{HgBa}_2\text{CaCu}_2\text{O}_6$, see Fig.~\ref{fig:HBCCO}(b)), the orbital spreads and $S$ change in opposite directions as a parameter of the method is varied, whether it is $\sigma$ in IWF, the outer energy window in MLWF, or $p_{\min}$ in PDWF. There is thus a trade-off between the localization of the Wannier functions and the fidelity of the downfolded bands, and in IWF it is controlled by the single dimensionless parameter $\sigma$.

\begin{table*}[htbp]
\centering
\begin{tabular}{lllccc}
\toprule
System & Method & Settings & $\Omega_\mathrm{D}$ (\AA$^2$) & Largest spread (\AA$^2$) & $S$ (eV) \\
\midrule
Si & IWF & $\sigma = 0.01$ & 0.5789 & 4.9121 & 0.0537 \\
 & & $\sigma = 0.05$ & 0.4946 & 4.5082 & 0.2111 \\
 & & $\sigma = 0.20$ & 0.3274 & 3.4976 & 0.6921 \\
\cmidrule(l){2-6}
 & SCDM & & 1.6590 & 7.2116 & 0.0581 \\
\cmidrule(l){2-6}
 & MLWF & & 0.1312 & 2.3940 & 0.6153 \\
\midrule
Graphene & IWF & $\sigma = 0.01$ & 0.4079 & 4.3683 & 0.0375 \\
 & & $\sigma = 0.05$ & 0.2518 & 3.8320 & 0.3487 \\
 & & $\sigma = 0.20$ & 0.0717 & 2.1862 & 0.8131 \\
\cmidrule(l){2-6}
 & SCDM & & 2.4256 & 6.0431 & 0.5089 \\
\cmidrule(l){2-6}
 & MLWF & & 0.0105 & 0.5969 & 1.0997 \\
\midrule
$\text{HgBa}_2\text{CaCu}_2\text{O}_6$ & IWF & $\sigma = 0.01$ & 2.4640 & 25.8337 & 0.2668 \\
 & & $\sigma = 0.05$ & 0.1815 & 8.5223 & 0.1178 \\
 & & $\sigma = 0.20$ & 0.0187 & 4.9308 & 0.2039 \\
\cmidrule(l){2-6}
 & MLWF & outer $[-7, 3]$ eV & $3.84\times 10^{-5}$ & 1.1140 & 0.6052 \\
 & & outer $[-10, 8]$ eV & $2.20\times 10^{-5}$ & 0.8311 & 1.4760 \\
 & & outer $[-10, 12]$ eV & $3.55\times 10^{-5}$ & 0.7327 & 1.9783 \\
\cmidrule(l){2-6}
 & PDWF & $p_\mathrm{min} = 0.01$ & 0.0126 & 1.7368 & 0.7109 \\
 & & $p_\mathrm{min} = 0.02$ & 0.0148 & 2.3427 & 0.4411 \\
 & & $p_\mathrm{min} = 0.05$ & 0.0179 & 4.3131 & 0.2043 \\
\bottomrule
\end{tabular}
\caption{Band-diagonal spread $\Omega_\mathrm{D}$, the largest individual spread $\max_n (\langle r^2\rangle_n - \bar{\mathbf{r}}_n^2)$ of the localized orbitals (in \AA$^2$), and the band energy spread $S$ (Eq.~\ref{eq:energy-spread}, in eV) for the different localization methods. $p_{\max} = 0.50$ for all PDWF calculations.}
\label{tab:spread_metrics}
\end{table*}

\section{Conclusions\label{sec:conclusion}}

We have shown above that the Intrinsic Wannier Function (IWF) method provides a simple way to downfold band structure given a target set of atomic characters. While the implementation here used an atomic orbital representation of the bands, the method can work with any computational basis, and the straightforward algorithm should enable it to be widely used in general materials electronic structure simulation. Further, the robustness of the procedure, and the dimensionless nature of the smearing parameter, mean that material-specific adjustments to the calculation are minimal. The algorithm is thus compatible with high throughput calculations~\cite{Vitale2020-ce}, a direction which will be interesting to explore in the future. 

\section*{Acknowledgements}

We thank Zhi-Hao Cui, Linqing Peng, Petra Shih, Tianyu Zhu and Yuhang Ai for providing enlightening suggestions for the conceptualization of this project. We also thank Arash Mostofi, Jaemo Lihm, Valerio Vitale, and Anil Damle for additional insights into SCDM. This work was supported by the US Department of Defense Multidisciplinary University Research Initiative (MURI) Grant No. W911NF2410292. 

~\\

\section*{Data Availability}
The source code is available in the Github repository https://github.com/LiShuoxue/iwf-preview.git. The calculation data is available in the Zenodo open repository (https://doi.org/10.5281/zenodo.24202342).

\bibliography{main.bib}

\onecolumngrid

\renewcommand{\thesubsection}{S\arabic{subsection}}
\renewcommand{\thefigure}{S\arabic{figure}}
\renewcommand{\thetable}{S\arabic{table}}
\renewcommand{\theequation}{S\arabic{equation}}

\setcounter{section}{0}
\setcounter{figure}{0}
\setcounter{table}{0}
\setcounter{equation}{0}

\section*{Supplementary Information}

\begin{figure*}[htbp]
\centering
\begin{minipage}[c]{0.50\textwidth}
  \centering
  \includegraphics[width=\linewidth]{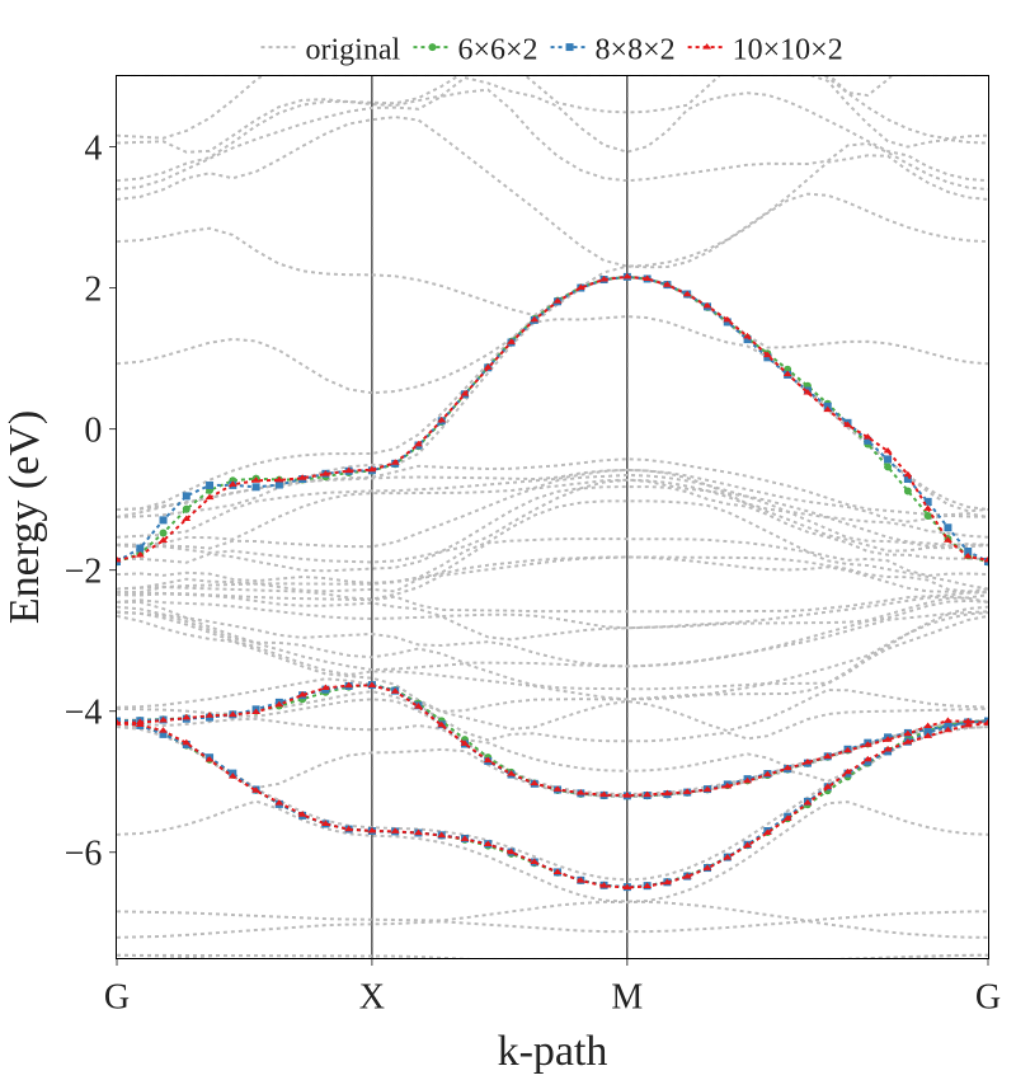}\\[-2pt]
  (a)
\end{minipage}\hfill
\begin{minipage}[c]{0.47\textwidth}
  \centering
  \small\setlength{\tabcolsep}{4pt}
  \begin{tabular}{lccc}
  \toprule
  $k$-mesh & $6\times 6\times 2$ & $8\times 8\times 2$ & $10\times 10\times 2$ \\
  $\varepsilon_d - \varepsilon_p$ (eV) & 1.705 & 1.708 & 1.727 \\
  $|t_{pd}|$ (eV) & 1.273 & 1.274 & 1.275 \\
  $|t_{pp}|$ (eV) & 0.745 & 0.745 & 0.742 \\
  $|t'_{pp}|$ (eV) & 0.011 & 0.017 & 0.015 \\
  $n_\mathrm{Cu}$ & 1.512 & 1.511 & 1.509 \\
  \bottomrule
  \end{tabular}\\[6pt]
  (b)
\end{minipage}
\caption{$k$-mesh convergence of the IWF ($\sigma = 0.05$) three-band model of $\text{HgBa}_2\text{CaCu}_2\text{O}_6$.
(a) Projected bands of the $6\times 6\times 2$, $8\times 8\times 2$ and $10\times 10\times 2$ meshes over the
ab initio bands (gray, $10\times 10\times 2$), all relative to the common Fermi level of the
$10\times 10\times 2$ mean field.
(b) The model parameters and the Cu population of the tight-binding
model (5 electrons per cell) for different $k$-meshes.}
\label{fig:hbcco_kconv}
\end{figure*}

\begin{table*}[h]
\centering
\begin{tabular}{llcccc}
\toprule
Method & Setting & $\varepsilon_d - \varepsilon_p$ & $|t_{pd}|$ & $|t_{pp}|$ & $|t'_{pp}|$ \\
\midrule
IWF & $\sigma = 0.05$ & 1.708 & 1.274 & 0.745 & 0.017 \\
 & $\sigma = 0.20$ & 1.680 & 1.276 & 0.749 & 0.025 \\
 \midrule
MLWF & outer $[-7, 3]$ eV & 1.831 & 1.216 & 0.661 & 0.166 \\
 & outer $[-10, 8]$ eV & 1.602 & 1.239 & 0.539 & 0.091 \\
 & outer $[-10, 12]$ eV & 1.387 & 1.257 & 0.437 & 0.019 \\
\midrule
PDWF ($p_\mathrm{max} = 0.50$) & $p_\mathrm{min} = 0.01$ & 2.094 & 1.240 & 0.636 & 0.133 \\
 & $p_\mathrm{min} = 0.02$ & 2.414 & 1.228 & 0.613 & 0.144 \\
  & $p_\mathrm{min} = 0.05$ & 2.932 & 1.176 & 0.543 & 0.186 \\
\bottomrule
\end{tabular}
\caption{Parameters of the three-band CuO$_2$-plane model (Cu $3d_{x^2-y^2}$, O $2p_x$, O $2p_y$) of $\text{HgBa}_2\text{CaCu}_2\text{O}_6$: the charge-transfer energy $\varepsilon_d - \varepsilon_p$, the
nearest-neighbor Cu--O hopping $t_{pd}$, the nearest-neighbor O--O hopping $t_{pp}$, and the
next-nearest-neighbor O--O hopping $t'_{pp}$. The hoppings are
averaged over the symmetry-equivalent bonds which agree within $2 \times 10^{-3}$ eV.}
\label{tab:hbcco_hoppings}
\end{table*}

\begin{table*}[h]
\centering
\begin{tabular}{llccc}
\toprule
Method & Setting & Cu--O & Cu--Cu & O--O \\
\midrule
IWF & $\sigma = 0.05$ & $-0.287$ & $-0.202$ & $-0.274$ \\
 & $\sigma = 0.20$ & $-0.282$ & $-0.218$ & $-0.314$ \\
\midrule
MLWF & outer $[-7, 3]$ eV & $-0.594$ & $-0.447$ & $-0.540$ \\
 & outer $[-10, 8]$ eV & $-0.460$ & $-0.380$ & $-0.435$ \\
 & outer $[-10, 12]$ eV & $-0.443$ & $-0.391$ & $-0.319$ \\
\midrule
PDWF ($p_\mathrm{max} = 0.50$) & $p_\mathrm{min} = 0.01$ & $-0.377$ & $-0.293$ & $-0.385$ \\
 & $p_\mathrm{min} = 0.02$ & $-0.391$ & $-0.312$ & $-0.356$ \\
 & $p_\mathrm{min} = 0.05$ & $-0.409$ & $-0.295$ & $-0.322$ \\
\bottomrule
\end{tabular}
\caption{Decay rates (in \AA$^{-1}$) of the real-space Hamiltonian $H_{ij}(\mathbf{R})$ of the
three-band CuO$_2$-plane models of $\text{HgBa}_2\text{CaCu}_2\text{O}_6$: slopes of the linear regression of $\ln|H_{ij}(\mathbf{R})|$
against the $i$--$j$ distance $|\mathbf{r}_j - \mathbf{r}_i - \mathbf{R}|$, taken as the minimum
distance under the $8\times 8\times 2$ supercell of the $k$-mesh, for the
Cu $3d_{x^2-y^2}$--O $2p$, Cu--Cu and O $2p$--O $2p$ channels. The on-site terms are excluded. The PDWF results use $p_\mathrm{max} = 0.50$ and no energy window.
A more negative slope indicates a faster decay.}
\label{tab:hbcco_hamiltonian_decay}
\end{table*}

\end{document}